\documentclass[sigconf,nonacm]{acmart}

\usepackage{xac}
\usepackage{array}
\usepackage{tcolorbox}
\usepackage{fvextra}

\AtBeginDocument{%
  }

\setcopyright{none}
\renewcommand\footnotetextcopyrightpermission[1]{}
\newcommand{\sys}{\textsf{\textsc{MindSpan}}\xspace}
\newcommand{\sysa}{\textsf{\textsc{Stitch}}\xspace}
\newcommand{\sysb}{\textsf{\textsc{\sys}}\xspace}

\newcommand{\uxd}[0]{\texttt{UXD}}

\AtBeginDocument{%
  \renewcommand{\pm}{\ifmmode\mbox{\texttt{PM}}\else\texttt{PM}\fi}%
}
\newcommand{\swe}[0]{\texttt{SWE}}
\newcommand{\uxr}[0]{\texttt{UXR}}

\definecolor{revPurple}{HTML}{6A3D9A}   %
\definecolor{revBlue}{HTML}{0B4F9F}     %
\definecolor{revTeal}{HTML}{008D91}     %
\definecolor{revGreen}{HTML}{2B7B2B}    %
\definecolor{revOrange}{HTML}{D95F02}   %

\newcommand{\rrone}[3]{\rribox{#1}{#2} {\color{revPurple}#3}}
\renewcommand{\rrone}[3]{{#3}}

\definecolor{litMagenta}{HTML}{B5179E}

\definecolor{roleBlue}{HTML}{0057B8}

\definecolor{timelessTeal}{HTML}{00766C}

\begin{document}

\title{Rethinking the UI of GenUI: A Tale of Two Designs}

\author{Xiang `Anthony' Chen}
\affiliation{%
  \institution{HCI Research, UCLA}
  \city{Los Angeles}
  \state{California}
  \country{United States}}

\author{Savvas Dimitrios Petridis}
\affiliation{%
  \institution{Google DeepMind}
  \city{New York}
  \state{New York}
  \country{United States}}

\author{Ruofei Du}
\email{ruofei@google.com}
\affiliation{%
  \institution{Google XR}
  \city{San Francisco}
  \state{California}
  \country{United States}}

\author{Yang Li}
\affiliation{%
  \institution{Google DeepMind}
  \city{Mountain View}
  \state{California}
  \country{United States}}

\renewcommand{\shortauthors}{Chen et al.}

\begin{abstract}
  Generative user interface (GenUI) tools use large models to create UI mockups from high-level descriptions; a recurring interaction paradigm starts with an unstructured prompt, followed by depth-first generation of a polished, high-fidelity prototype. To examine how it supports early-stage, 0-to-1 design exploration, we contrast this paradigm with structured input, breadth-first workflow, and low-fidelity generation. In our study, 24 industry professionals in early-stage product and UX design used a commercial GenUI system and our research prototype as study-era instances of each paradigm for a comparative evaluation. Findings show that 1) structured input surfaces key facets but raises entry barriers; 2) breadth-first workflow reveals more possibilities, but previewing multi-screen UX ideas remains hard; and 3) despite low fidelity's value, participants rated high fidelity as better suited to their design stage and to communicating ideas. Our findings contribute a model-agnostic, design-level account of how GenUI users specify intent, traverse alternatives, and interpret fidelity.
\end{abstract}
\begin{teaserfigure}
    \centering
    \includegraphics[width=\textwidth]{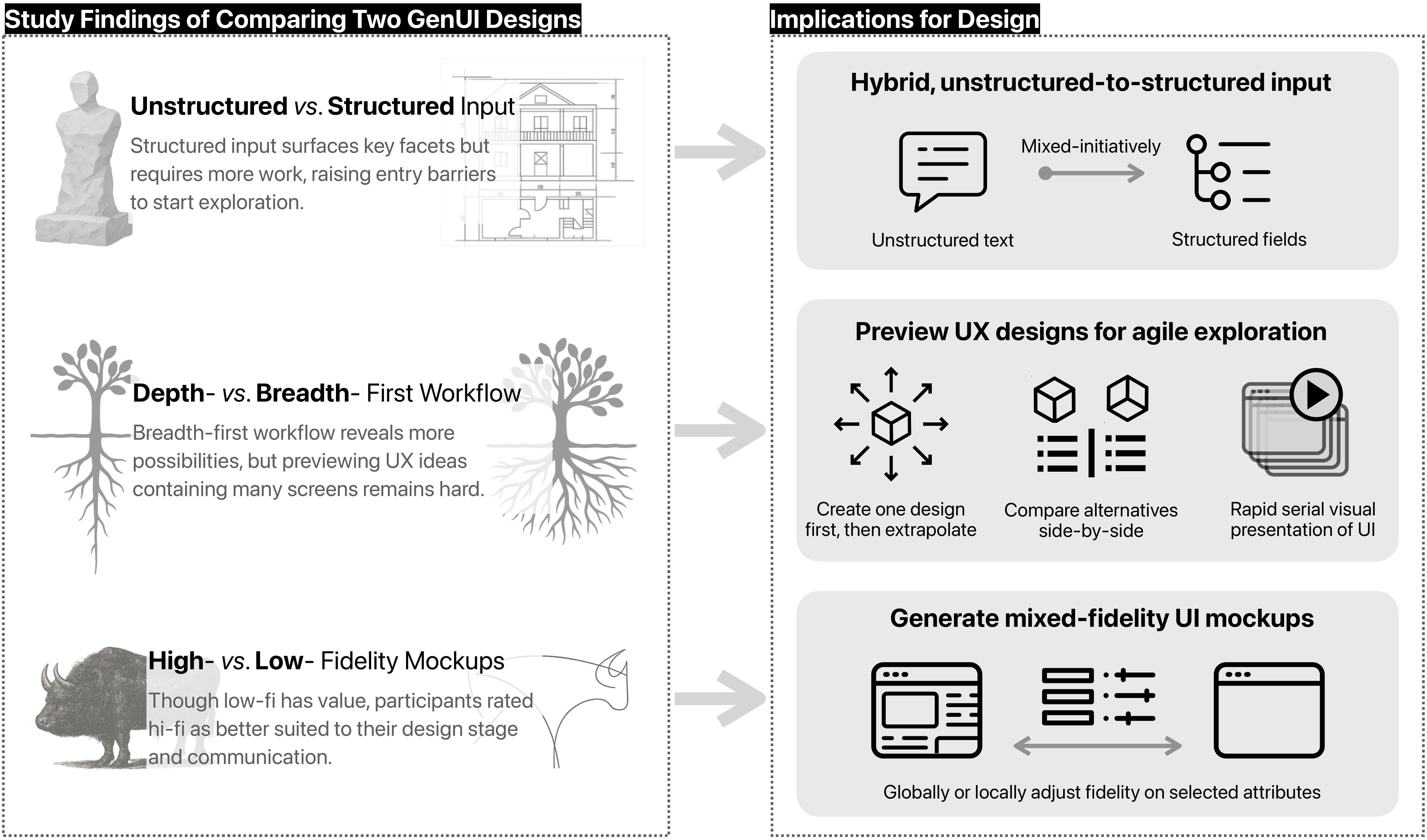}
    \caption{
      We studied two contrastive designs of GenUI---tools that employ large models to generate UI mockups---to understand their trade-offs and how tool-builders can calibrate design choices:
      \one Input: a conventional open-ended, unstructured prompt \vs a structured form that guides users to provide key information;
      \two Workflow: delving into a design right away (depth-first) \vs starting with exploring a design space (breadth-first);
      \three Output: generating polished, high-fidelity \vs rough, low-fidelity UI mockups.
    }
    \label{fg:teaser}
\end{teaserfigure}



\maketitle

\section{Introduction}

Generative AI has made polished, high-fidelity prototypes cheap to produce, prompting calls to revisit foundational design wisdom about when fidelity helps and when it hinders \cite{hundhausen2026fidelity}. At the center of this shift are \textbf{Generative User Interfaces (GenUI)}---UI screen mockups generated by pre-trained large models from high-level inputs (\eg a textual description)---with the potential to broaden access to UX design exploration.
One major promise of GenUI is supporting the important early-stage, ``0-to-1''\footnote{There are two distinct types of design needs that GenUI tools can support:
``0-to-1'' \vs ``n-to-(n+1)'', where the former is about exploring a new feature or product from scratch and the latter is about adding or changing features on an existing product.} ideation \cite{10.1145/3715336.3735780},
which is often cross-functional: stakeholders with different professional responsibilities may use prototypes to articulate requirements, explore user journeys, and develop design ideas \cite{10.1007/11774129_15, MELEGATI2024107516, banfield2015design}. We therefore focus on professionals who are substantively involved in this early phase of product and UX design, without treating professional role as the phenomenon under study.

Empirical studies document the consequences of cheap fidelity for design practice. Practitioners across technical skill levels have come to produce high-fidelity prototypes as a matter of course, expecting to carry forward artifacts that were once disposable \cite{kobiella2026throwaway}; product teams have reorganized prototyping around speed, often at the expense of reflection \cite{li2026vibeteams}; and AI-generated prototypes can match human-made ones on usability while scoring lower on originality \cite{romero2026usable}.
GenUI tools embody this shift. Several of them%
\rrone{a}{R1}%
{---\eg Uizard~\cite{uizard}, Visily~\cite{visily}, Luny AI~\cite{luny}, Figma Make~\cite{figma_make}, and Stitch~\cite{stitch}---}%
instantiated a recurring interaction paradigm inherited from chat-based AI: users began with unstructured text prompts, and the system pursued a ``depth-first'' workflow by generating a complete high-fidelity UI mock-up (\eg in HTML/CSS) before accepting further refinements.

This interaction paradigm offers speed and directness, but it also creates trade-offs that do not depend on a particular product or model version.
First, an unstructured text prompt allows for easy and free expression, yet information provided by users might be incomplete and ambiguous \cite{10.1145/3586183.3606725, morris2024prompting, 10.1145/3613904.3642400, 10.1145/3706598.3714259}, leading to the model's suboptimal interpretation of their design needs.
Prior work frames this as a \textit{gulf of envisioning} specific to UI generation, where users struggle both to articulate design intent and to trace how that intent surfaces in the generated result \cite{park2026bridging, xu2026ooprompt}.
Second, going straight from prompts to UI generation presents a fast result but might overlook the discover phase \cite{designcouncil_framework_innovation}---divergent \cite{goldschmidt2016linkographic} exploration of the underlying design space \cite{10.1145/3613904.3642400} as an important purpose of prototyping \cite{368126,houde1997prototypes,10.1145/1978942.1979359,10.1145/1124772.1124960,10.1145/1375761.1375762}.
Third, generating a high-fidelity UI mock-up matches what conventional design tools (\eg Figma) can offer, yet its attention to fine details risks prematurely locking the design into a specific direction \cite{doi:10.1177/154193128903300405,10.1145/223904.223910,buxton2010sketching, 10.1145/1357054.1357074}---a risk that grows as high fidelity becomes the inexpensive default.
Prior GenUI research has explored a range of systems \cite{10.1145/3706598.3713924, 10.1145/3746059.3747672, 10.1145/3773035, 11024434}---including mechanisms to steer generation via semantics, visual flow, or learned preferences \cite{park2026bridging, garg2026controlgui, liu2026alignui}---yet a systematic investigation of these three design choices remains absent from the literature.

In this research, we examine the design of GenUI by juxtaposing the unstructured, depth-first, high-fidelity paradigm against a contrastive approach along three dimensions that span the end-to-end interaction:
\begin{itemize} [leftmargin=0.25in]
    \item Input: Instead of unstructured text prompts, we build on prior work \cite{fi16110406, LU2024102835, 10.1145/3706598.3713785} and
    instrument a \textit{structured} input format that employs the critical user journey (CUJ) method \cite{vachon2025security},
    guiding users to specify the context, user, goal, and tasks related to their design needs.
    \item Workflow: Instead of going depth-first, we employ a \textit{breadth-first} workflow, found in numerous creativity support tools \cite{10.1145/1124772.1124960,10.1145/1375761.1375762, 10.1145/3613904.3642400, 10.1145/3173574.3174025, 10.1145/3706598.3714115}, that starts with generating a design space for users to see and explore multiple design alternatives before delving into one to generate the UI mockups.
    \item Output: Instead of high fidelity, we opt to generate \textit{low-fidelity} UI mockups that are less polished, yet a lack of fine details could in turn reduce information load while allowing more focus on the high-level design ideas.
\end{itemize}

\textbf{Our central research question} is: \textit{How do these design choices of input, workflow, and output affect the usage patterns, benefits, and trade-offs of GenUI tools for early-stage design exploration?}

To investigate this question along the three dimensions, we created \sys---a proof-of-concept GenUI tool that implemented the core design ideas of structured input, breadth-first workflow, and low-fidelity output.
Using \sys as a probe, we then conducted hour-long workshops with 24 industry professionals involved in early-stage product and UX design from the same software company.
We recruited both UX designers and product managers, who bring two different but closely related perspectives to early-stage design; studying the differences between these roles is beyond the scope of this work.
Following a within-subject design, each participant used two GenUI tools: our prototype and the version of Stitch \cite{stitch} used in the study, which instantiated the unstructured, depth-first, high-fidelity paradigm.
After each task, participants filled out survey questions to rate the two GenUI designs' input, workflow, and output.
We use the system names to document the apparatus; our analysis treats them as probes of contrasting interaction paradigms, not as enduring performance baselines or a ranking of products and models.
Within two workdays after the workshop, we conducted one-on-one semi-structured interviews with each participant to further discuss how they compared the two GenUI tool designs.

Our key findings are as follows:
\begin{itemize} [leftmargin=0.25in]
    \item While structured input helped participants think through the many key facets of UX design (user, goal, and tasks), it also increased the barrier to entry compared to unstructured input, calling for a hybrid approach and more support for a ``warm start'', especially providing examples to build off of.
    \item The breadth-first workflow outperformed the depth-first approach in challenging participants' assumptions and led them to discover otherwise overlooked design possibilities. However, one main challenge is previewing a UX design idea when exploring many, as each idea itself also consisted of multiple screens of information to support multiple task flows.
    \item Despite the reasons to favor low fidelity for design exploration, participants rated high fidelity as better matching their design stage and their need to communicate ideas, and many described it as a better fit for their professional practices. Prior exposure to product-like AI prototyping (including ``vibe coding'') and expectations built during breadth-first exploration possibly amplified disappointment with low-fidelity mockups.
\end{itemize}

Building off of these findings, we discuss implications for designing GenUI and other AI-powered creativity support tools (illustrated in \fgref{teaser}), including
\one A hybrid, mixed-initiative, unstructured-to-structure input to ease into GenUI exploration.
\two Previews of UX designs to support agile exploration, \eg create one design first, then extrapolate; compare alternatives side-by-side; and rapid serial visual presentations of UI screens.
\three Supporting mixed-fidelity generation so users can choose or adjust fidelity globally or locally according to their design purpose and stage.

\textbf{The main contribution} is an empirical, design-level account of the trade-offs among input structure, exploration workflow, and output fidelity, examined through two concrete system configurations. The contribution is not a benchmark of the particular systems used in the study; it is a set of findings and design implications about choices that continue to organize GenUI interaction even as underlying models and products change.

\section{Related Work}

GenUI employs AI to create prototypes (UI mockups) that support UI/UX work, which leads to our review of two related bodies of literature: the use of prototypes in UI/UX design process (\secref{uiux}) and studies of how designers use AI in their work (\secref{use_ai}).
Then, we focus on one specific type of AI for UX work---GenUI tools---and review its technical foundations and recurring interaction paradigms.

\subsection{Prototypes in UI/UX Design Process}
\label{sec:uiux}

The use of prototypes is a long-standing practice in the software industry that predated modern UI/UX design process.
Lichter \etal studied how industrial software teams adopted prototyping practices
\cite{368126} and
their findings surface three key attributes of prototypes, that
\one prototyping is part of the \textit{evolutionary} approach of software development;
\two prototyping aims at creating \textit{early working versions} of a system or application; and 
\three prototyping provides a medium for \textit{communication} among team members as well as users.
Later work on UI/UX design process echoed and built on these three characteristics.
Houde and Hill summarized three specific prototyping purposes: understanding the \textit{role} the system plays in its context of usage, exploring how the system looks and how users interact with it (\ie \textit{look and feel}), and probing technical solutions to \textit{implement} the system \cite{houde1997prototypes}.
Lim \etal proposed an anatomy of prototypes comprised of two dimensions: \textit{filtering} certain aspects of the system design and \textit{manifesting} the prototype in a certain medium, which enables designers to take a systemic view of prototypes and prototyping practices \cite{10.1145/1375761.1375762}.
Among all the above studies, one shared sentiment is that prototypes---whether created by humans or AI---
should facilitate the iterative design and development of products, from exploring early ideas, to experimenting with implementation solutions, and to evolving an early working version to a product-ready artifact \cite{368126,houde1997prototypes,10.1145/1375761.1375762}.
Our study is in part motivated by these findings, as we hypothesized that a breadth-first workflow and a low-fidelity output can facilitate the use of generated UI mockups to drive the iterative UI/UX process.
Notably, this body of prototyping wisdom was established when fidelity was expensive---a premise that generative AI has unsettled (\secref{lit-fidelity}).

\subsection{Studies of Designers' Use of GenAI}
\label{sec:use_ai}
Research on AI in UX design has examined its integration into designers' workflows \cite{lu_ai_2024, 10.1145/3764687.3764719} (\eg the double-diamond model \cite{designcouncil_framework_innovation}), its capacity to contextualize user data \cite{10.1145/3295750.3298942}, and recurring patterns and challenges in adoption \cite{10.1145/3544549.3573874, 10.1145/3613905.3650878, stige_artificial_2023}. GenAI systems have also supported activities such as generating personas and constrained design examples \cite{10.1145/3706598.3713785} and mapping user journeys through inspirations and progressive discussions \cite{10.1145/3706598.3713479}.
The GenUI tool we studied (\sys) similarly leverages GenAI to generate user personas and journeys, but extends this by integrating user-centered information to create UI mockups.

Empirical work shows that GenAI offers opportunities for UX practice---such as generating personas, illustrating journeys \cite{uusitalo_clay_2024}, and fostering divergent thinking through design exploration \cite{10.1145/3706598.3713500}---while also introducing persistent challenges. These tensions span the end-to-end design process and include a lack of shared team practices and training \cite{takaffoli_generative_2024}, difficulties in hand-offs and cross-role communication \cite{li_user_2024}, systemic collaborative hurdles \cite{subramonyam2024content}, and skepticism from designers about GenAI's promised potential \cite{conf/chi/Naqvi}.
The rise of ``vibe coding'' \cite{sarkar2025vibe} has further reconfigured this landscape.
Studying product teams, Li \etal reported that vibe coding introduces new asymmetries of trust and responsibility across roles \cite{li2026vibeteams}; Pimenova \etal similarly characterized it as a co-constituted practice that favors flow and experimentation over upfront specification \cite{pimenova2025goodvibrations}.
These shifts blur the boundary between designers and developers in UI prototyping \cite{zhang2026genvibe}, a boundary that GenUI tools occupy.
Taken together, this body of work documents particular technologies and practices at specific points in time. Our research instead organizes the inquiry around three persistent GenUI design dimensions, using a contrastive approach to surface the benefits and trade-offs of alternative choices along each dimension.

Other research focused on studying GenAI's integration in UX design tools. Feng \etal developed a Figma plug-in to examine how LLMs translate requirements into concrete design artifacts \cite{feng2024canvil}, and PromptInfuser links UI elements in Figma to model inputs and outputs to let designers simulate and probe AI functionality at design time \cite{10.1145/3544549.3585628}. Closest to our work, Chen \etal \cite{10.1145/3715336.3735780} studied multiple design and non-design roles using one commercial GenUI tool, reporting how GenUI supports each role, along with challenges and improvement opportunities. While \cite{10.1145/3715336.3735780} addressed a broad range of GenUI issues, our studies delve deeper into three key aspects (input, workflow, and output) and are grounded in a contrastive study of two different GenUI designs rather than a single tool.

\subsection{Overview of GenUI Tools}

There are three families of GenUI tools based on different UI representations: 
\one generating pixels of a UI screen image (\eg using Generative Adversarial Networks \cite{goodfellow2014generative} or Diffusion models \cite{JMLR:v23:21-0635} to provide design examples \cite{10.1145/3491102.3517511});
\two generating hierarchical data structure (\eg transforming an initial set of UI elements into a refined layout \cite{lilayoutgan}), and;
\three generating source code (\eg using programming-by-demonstration to implement new interactive features \cite{conf/uist/SarmahDWLLC20} or transformers to generate tokens representing and rendering a UI \cite{huang2021creating, wu-2024-uicoder}).

Code-generation approaches have been incorporated into commercial tools such as Uizard~\cite{uizard}, Visily~\cite{visily}, UX Pilot~\cite{uxpilot}, Luny AI~\cite{luny}, Figma Make~\cite{figma_make}, Emergent~\cite{emergent}, Banani~\cite{banani}, Framer~\cite{framer}, and Google's Stitch~\cite{stitch}.
Across this set, tools including Uizard, Stitch, Visily, and Luny AI instantiated the same combination of design choices: unstructured inputs (natural language, sketches, or screenshots), a depth-first generation strategy, and high-fidelity polished mock-ups produced in a single pass.
For example, Uizard~\cite{uizard} utilizes a transformer-based heuristic to map natural language tokens to a predefined design system schema, automatically generating layout structures, color variables, and font pairings that align with the semantic intent of the prompt.
The engine executes a high-fidelity render by bypassing low-fidelity wireframing, directly instantiating complex UI components and multi-screen navigation flows from a single string of unstructured text.
Notably, a few exceptions do depart from this design paradigm: UX Pilot introduces more structured input via describing individual screens, while Div-idy \cite{dividy} emphasizes breadth-first exploration by decomposing prompts into modular components and reassembling them via a multi-stage pipeline.
Leviathan \etal showed that, given appropriate prompting and tools, the LMs they evaluated could generate custom interfaces that human raters preferred over conventional text output and judged comparable to expert-crafted UIs in roughly half of cases \cite{leviathan2026genui}.
Such capability gains, however, do not by themselves resolve how well a generated design serves a designer's intent.
Imteyaz \etal audited five GenUI tools and found ``design theater''---over a quarter of the design rationales these tools present to users are not actually realized in the generated interface, and the tools recognize only about half of the UX principles embedded in a task \cite{imteyaz2026designtheater}.
Research systems address parts of this gap by introducing controllable intermediate structure: semantic guidance that lets users specify and trace design semantics \cite{park2026bridging}, perceptual visual flow and wireframes as partial specifications for rapid GUI exploration \cite{garg2026controlgui}, and preference-aligned reasoning over UI controls \cite{liu2026alignui}. Our work is complementary: rather than benchmark generation quality, we use two study-era systems to contrast end-to-end interaction paradigms and treat their underlying models as part of the apparatus.

\section{GenUI's Three Design Dimensions}
\label{sec:lit-three-aspects}

To contextualize our co-design and comparison studies, we outline three key design dimensions of GenUI.

In contrast to general guidelines for designing human-AI collaboration \cite{horvitz1999principles, 10.1145/3290605.3300233, 10.1145/3334480.3381069} (\eg when to invoke AI and recover from its error), the following dimensions are grounded in the end-to-end process of interacting with GenUI, specifically addressing what input users should provide and how, whether the workflow should delve into one design or explore breadths first, and which fidelity is most appropriate as output.

\subsection{Unstructured \vs Structured Input}
HCI research has long contemplated the benefits and trade-offs between unstructured and structured input, from early work of a structural analysis on input devices \cite{10.1145/123078.128726} and subtasks \cite{foley1980human} to research that envisaged unstructured input manifested as natural language interaction \cite{10.1145/800250.807503}.

This unstructured, conversational approach became a hallmark of mixed-initiative human-AI collaboration \cite{horvitz1999principles} and was widely adopted in LM-based interfaces.

Unstructured prompting, commonly found in chat-based AI interfaces, works well for non-technical users, as it is highly expressive and requires no learning curve \cite{10855599, morris2024prompting}. LMs can respond to prompts by producing numerous outputs instantly, sometimes leading to serendipitous discoveries \cite{10.1145/3586183.3606725, 10.1145/3613904.3642400}. However, crafting effective prompts to accurately capture creative intent is difficult \cite{10.1145/3586183.3606725, 10.1145/3706598.3714259, 10855599}, often involving laborious trial-and-error \cite{10.1145/3586183.3606725} and a high cognitive load for non-experts who may lack the necessary vocabulary or domain knowledge \cite{10.1145/3586183.3606725, 10.1145/3706598.3714259, 10855599, morris2024prompting}. This often leads to premature convergence on limited ideas (fixation), preventing users from exploring the vast design space in a structured way \cite{10.1145/3613904.3642400}.

In response, a line of work reintroduces structure into prompting.
Park \etal derived a hierarchy of design semantics from UI prompting guidelines and let users specify, visualize, and trace those semantics in the generated UI, which improved users' perceived control over expressing intent and interpreting outcomes \cite{park2026bridging}.
Xu \etal took a more general approach, reifying intents into structured, reusable artifacts so that prompting becomes modular and iterative rather than a monolithic string \cite{xu2026ooprompt}.
The structured input we study is in the same spirit but is scoped to UX design specifically, using the critical user journey as its organizing schema.

\subsection{Depth- \vs Breadth-First Workflow}
There has been a long-standing discussion on the tension between going deep \vs broad.
Shneiderman's mantra advocated starting with an overview (breadth) then filter and zoom in to examine details (depth) \cite{SHNEIDERMAN2003364}. Design literature found that divergent (breadth-first) and convergent (depth-first) activities are both key to the creative process and often occur in cycles \cite{goldschmidt2016linkographic}.
Among creativity support tools, exposing users to a breadth of design possibilities has been a hallmark of the workflow \cite{10.1145/1978942.1979359}, either by sampling the underlying parameter space to show design examples \cite{10.1145/571985.571996, conf/uist/ChenKMGCH16}, by clustering similar designs to provide an overview \cite{10.1145/3173574.3174025, 10.1145/3490099.3511156}, or by formulating high-level design dimensions to support systemic exploration \cite{10.1145/3613904.3642400, 10.1145/3706598.3714115}.
Specifically related to GenUI, prior work considered UI prototypes as a means for ``traversing a design space'' \cite{10.1145/1375761.1375762} and argued for exploring many than just a few \cite{10.1145/1124772.1124960, buxton2010sketching}.
When generating a specific UI design, the use of intermediate representation (\eg editable UI components~\cite{10.1145/3746059.3747672, 10.1145/3773035}) is another way of allowing designers to explore and control variations before converging to a specific version.

Research has extended breadth-first exploration to generative settings.
Zamfirescu-Pereira \etal built an IDE that surfaces alternative problem framings alongside alternative solutions, letting programmers parallelize design phases and cover a broader design space than code generation alone affords \cite{zamfirescu2025beyondcode}.
For GUIs specifically, Garg \etal argued that generative tools fail early-stage exploration precisely because they demand more specification detail than a designer has at that point, and proposed under-specified inputs (prompts, wireframes, visual flows) that yield many alternatives cheaply \cite{garg2026controlgui}.
A parallel concern is that naively sampling an LM many times returns candidates clustered in the same semantic basin, so breadth must be engineered rather than assumed \cite{ibrahim2026anchorless}.
Relevant to a challenge we surface later, Lu \etal noted that AI-powered UX tools focus on individual static screens and overlook experiences that span multiple screens, and offered pattern annotation over multi-screen user flows as one remedy \cite{lu2024flowy}.

\subsection{High- \vs Low-Fidelity Generation}
\label{sec:lit-fidelity}
Literature has shown that both low and high-fidelity prototypes have distinct roles and benefits depending heavily on the specific stage and goal of design. Low-fidelity prototypes, characterized by limited functionality and interaction, are valuable in the early stages for quickly generating and exploring a breadth of ideas \cite{buxton2010sketching, 10.1145/1357054.1357074, 10.1145/223904.223910, 10.1145/223500.223514}.
They allow designers to focus on high-level structure and behavior, facilitating ``getting the right design'' and avoiding premature fixation on details that can stifle creativity \cite{10.1145/1357054.1357074, 10.1145/223904.223910, 10.1145/223500.223514, doi:10.1177/154193120204600513}. Conversely, high-fidelity prototypes offer full interactivity and a realistic appearance, making them suitable for later stages to ``get the design right'' \cite{10.1145/238386.238516, 10.1145/223500.223514}.
Premature application of high-fidelity methods can be harmful by quashing innovative ideas, misdirecting focus to minor issues, or failing to account for cultural adoption over time \cite{10.1145/1357054.1357074}.

Generative AI unsettles this literature's central premise that fidelity is costly to produce \cite{hundhausen2026fidelity, kobiella2026throwaway}.
At the same time, cheap high fidelity is not neutral: in a blind comparison of 92 participants, AI-generated prototypes matched human-made ones on usability and efficiency but scored lower on originality and novelty, suggesting that generated polish comes with conventionalized structure \cite{romero2026usable}.
Whether professionals still see value in low fidelity under these conditions is an open question that our study takes up empirically.

To summarize, input structure, exploration workflow, and output fidelity provide a model- and product-agnostic way to characterize GenUI interaction, yet the literature has largely examined them in isolation. We address this gap by comparing their trade-offs as manifested in two concrete GenUI systems.

\section{A Tale of Two GenUI Designs}
\label{sec:design}

In this section, we describe the two study configurations as concrete instantiations of contrasting GenUI designs. The version of \sysa used in the study employed unstructured input and a depth-first workflow to generate high-fidelity UI mockups. Our research prototype, \sys (implementation details in \secref{implementation}), used structured input and a breadth-first workflow to generate low-fidelity UI mockups. The comparison concerns these design configurations rather than capabilities permanently associated with either system name.

In the scope of this work, we consider GenUI's target users to be professionals substantively involved in early product and UX design, which is often cross-functional \cite{10.1007/11774129_15, MELEGATI2024107516}.
To construct our narrative, we will follow Larry, a product manager (\pm) with a background in design, as he uses these two different GenUI tools (dubbed \sysa and \sysb) to explore ideas for features in a new software product---a Web app that helps busy parents plan weekend activities with their children.
Below we will juxtapose and contrast how Larry performs each key step differently using the two GenUI tools.

\subsection{Unstructured \vs Structured Input}

\paragraph{\sysa}
As shown in \fgref{genui_designs}, after Larry starts a new project, he writes a single plain-English prompt just like how he uses other conversational large models:
``\textit{Design a web app that helps busy parents plan weekend activities with their children}''.
Then he feels like adding a little more detail, so he continues typing ``\textit{... Maybe include features like Discover, Activity Details, Plan/Itinerary, Map, and Share}''.

\paragraph{\sysb}
In contrast, Larry needs to fill in a structured input form, including the four required CUJ-based input fields---context, user, goal, and tasks, where the default placeholder text explains what each field is for. 
Larry is not prepared to have all this information ready, so he pauses to think about each field, \eg the goal of the parent using this app, which should go beyond simply keeping their children busy during the weekend.
Larry considers a few possible goals, \eg to have fun, to stay connected as a family, and to educate the children.
He decides that he wants to explore a combination of fun and educational activities and fills in the goal accordingly.
Then, Larry also needs to fill in the tasks---what the user needs to do to achieve the goal, which forces him to think a little more concretely. Larry believes that the main task should be creating a plan for the upcoming weekend and he creates that one task accordingly.

\begin{figure*}[t]
\centering
\includegraphics[width=1.0\textwidth]{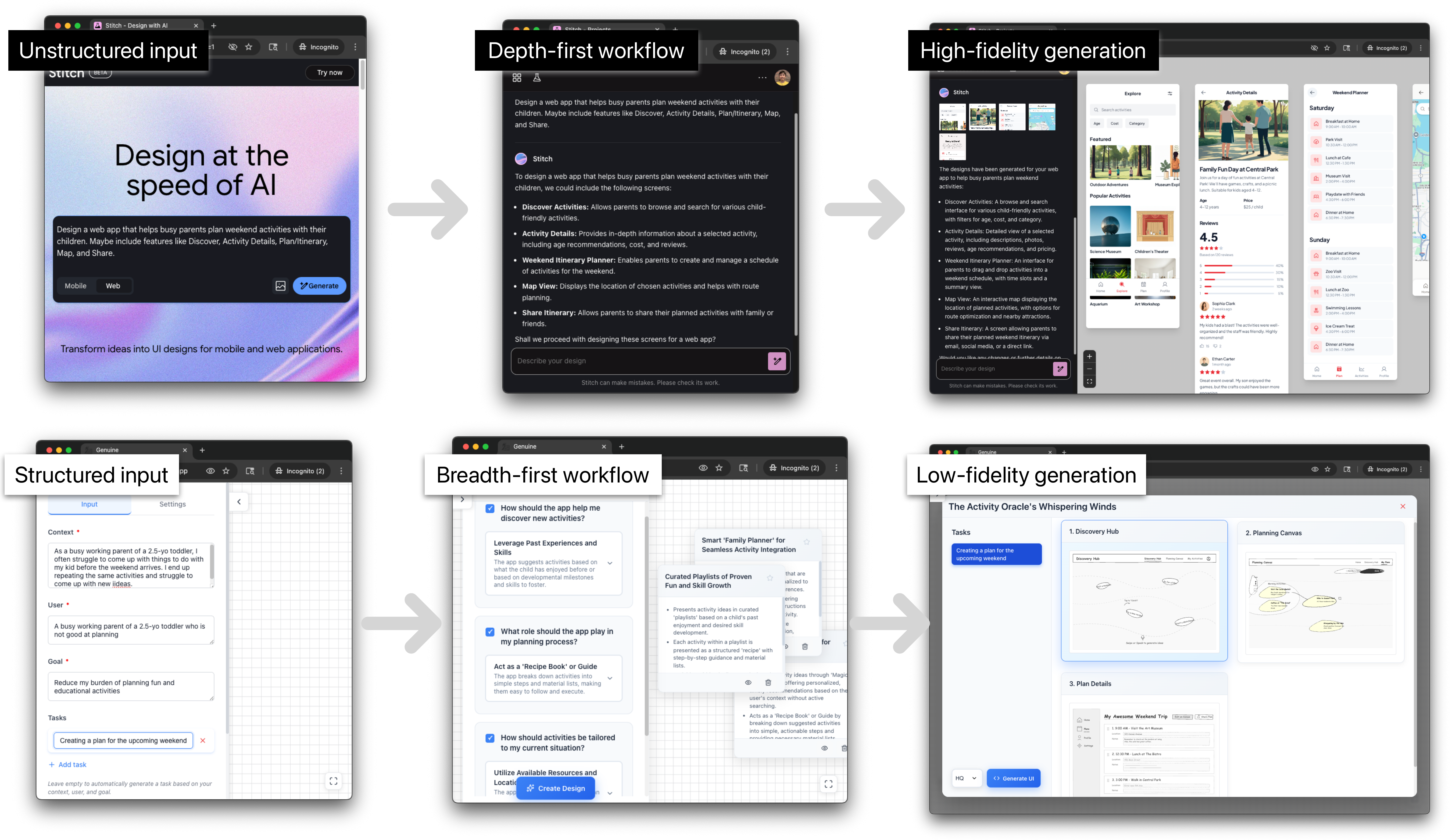}
\caption{
Comparison of two GenUI designs.
Top: based on Stitch \cite{stitch}, starts with an unstructured prompt and delves depth-first into the generation of a high-fidelity UI mockup, whereas Bottom: our contrastive design, starts with a structured input and a breadth-first exploration of the design space, and then generates low-fidelity UI mockups.
}
\label{fg:genui_designs}
\end{figure*}

\subsection{Depth- \vs Breadth-First Workflow}

\paragraph{\sysa}
As shown in \fgref{genui_designs}, after entering the prompt, Larry then hits a button that returns a list of screen descriptions in text and asks if he wants to proceed and generate those screens for an app mockup.
Glancing at the bullets and seeing that the descriptions do include what he mentioned earlier in the prompt (\eg Discovery Activities, Activities Details), Larry gives \sysa the go-ahead to generate the UI mockups.

\paragraph{\sysb}
After entering the CUJ information, \sys first creates a design space that consists of multiple dimensions, each phrased as a question calling for Larry to ponder a key design decision, \eg ``What role should the app play in my planning process?''
Larry can see and select options to respond to each dimension's question: each option leads with a short summative phrase (\eg ``Act as a `Recipe Book' or Guide''), followed by a brief explanatory sentence.
Larry can also uncheck a dimension if it feels less relevant.
\sys persistently shows the design space as a panel on the left side of the screen, allowing Larry to revisit it later to explore other different designs.

Next, Larry clicks the ``Create Design'' button to instantiate a specific idea based on the selected options in the design space.
\sys shows a design as a sticky-note-like card that summarizes the high-level concept as bullet points, 
placed on a canvas that supports direct manipulation, \eg separating designs into groups, or arranging them close to one another for comparison.
Larry can also star a design card to mark it as a favorite, or delete it.

\subsection{High- \vs Low-Fidelity UI Generation}

\paragraph{\sysa}
As shown in \fgref{genui_designs}, \sysa generates high-fidelity screen mockups with polished typography, spacing, and components.
Larry clicks into each generated screen and begins editing in place using the customization panel (\eg changing colors) or back to the conversational panel via interactive chat---for example:
``\textit{On Discover, add filter chips for ages 2–4 / 5–7 / 8–10 / 11–13; add Free and Indoor toggles}''
and
``\textit{On Map, add route preview for the current itinerary}''.

\paragraph{\sysb}
Larry selects one design card that interests him and opens its UI view, which first shows how the task---specified earlier in the structured input---corresponds to a flow of screens, each described in a one-liner text (no mockups generated yet).
Intrigued by the descriptions, Larry proceeds to generate UI codes for the screens using a default fast mode, which enables a faster (albeit rougher) ``preview'' version of the design\footnote{A high-quality (HQ) but slower mode is also available}.
As shown in \fgref{genui_designs}, the generated UI codes render each screen as a wireframe-like low-fidelity prototype.
Larry opens the full view of a screen, navigates to the previous or next one in the task flow, and reads about how a user would interact with each screen to perform the task.
Then, he closes the UI view and goes back to the deck of design cards.

\section{Implementation Overview}
\label{sec:implementation}

We describe a high-level overview (\fgref{system}) of \sys's implementation, while leaving low-level details (\eg step-by-step prompts in \sys) in the Appendix~\ref{sec:implementation_details}.

We implemented the front-end of \sys as a Web app using Vite \cite{vite} with React \cite{react2025}. 
During the study deployment, the back-end routed different prompt stages among three Gemini models \cite{google_gemini_models}: \textsf{2.5-pro}, \textsf{2.5-flash}, and \textsf{2.5-flash-lite}. We report these model names to document the apparatus, not because the contribution depends on these versions remaining current or best-performing. The interaction and orchestration logic was separated from the model calls; routing among the study-era models primarily balanced output quality and latency. The Appendix reports the exact model routing and prompts used for reproducibility.

\begin{figure*}[t]
\centering
\includegraphics[width=1\textwidth]{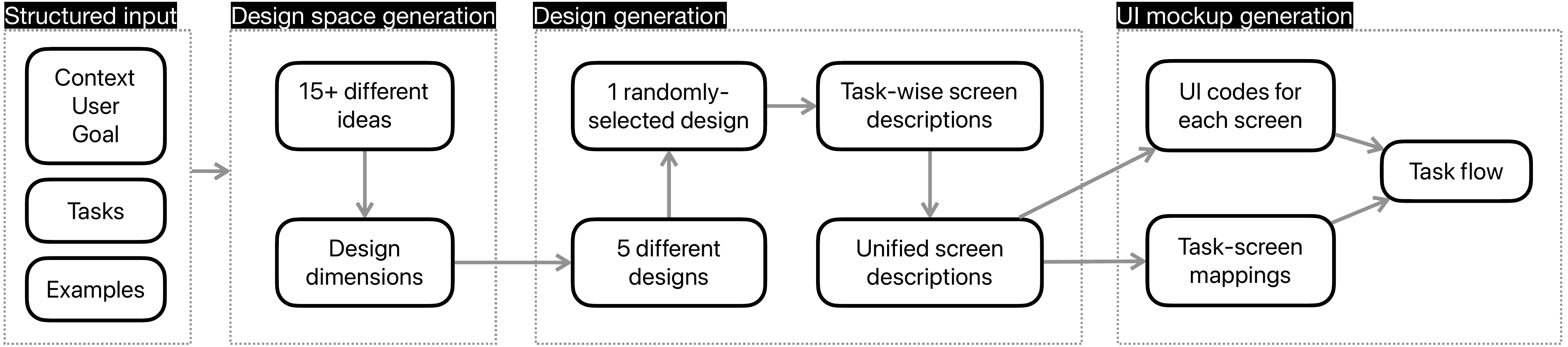}
\caption{\textbf{Block diagram of \sys's implementation.}}
\label{fg:system}
\end{figure*}

First, to create a design space, we employ a chain of two steps.
Given all the provided input (\eg context and CUJ information), 

we prompt the model to \textit{brainstorm $n$ different ideas}.
We chose $n=15$ to strike a balance between sufficient diversity and preventing additional latency.

Then, 
we prompt the model to analyze the ideas to \textit{identify common design dimensions} and create multiple options to choose from for each dimension.

To create a specific design, we prompt the model to implement the selected options in the design space and 

\textit{generate $n$ distinct high-level design concepts}, from which we \textit{randomly select one} to show to the user.
We chose $n=5$ for a similar reason of balancing diversity and latency.

Each model-generated design consists of a concise description and a detailed description framed in a way that a UX designer can understand and implement the concept into specific UI screens.
Next, we prompt the model to build on the generated design and create textual descriptions of each UI screen.
Given that there might be multiple tasks the UI screens need to support, we first prompt the model to generate \textit{task-wise screen descriptions}---that is, each task has its own sequence of screens, and then combine them into \textit{a unified sequence that supports all tasks}.

Then, we prompt the model to generate \textit{task-screen mapping}---to look back at each individual task and identify which screen(s) support that task, creating a mapping between each task and the corresponding screens.

To \textit{generate UI code for a screen}, we prompt the model with the screen's textual description to generate SVG code that renders a low-fidelity wireframe-like mockup.
We choose SVG over conventional HTML because its shape-based rendering naturally aligns with the look and feel of wireframes, while HTML's layout system tends to resemble polished, high-fidelity screens.
Finally, to \textit{generate a task flow}, we use the task-screen mapping to retrieve its screens' UI codes. For each screen involved in this task, we prompt the model to examine its key interaction (generated as part of the screen description) in the context of the generated UI codes and to describe the user interaction---how a target user might interact with the screen.
We present the user interaction description alongside the rendering of each screen's UI code, thus presenting how the user progresses in the task on this screen.

\section{Comparison Study}

We conducted a series of hands-on sessions to study how the two designs of GenUI tool support users' 0-to-1 exploratory design tasks, focused on juxtaposing their benefits and trade-offs.

We hypothesize that---
\begin{itemize} [leftmargin=0.25in]
    \item Input: Compared to an unstructured prompt, users will benefit more from structured input in breaking down a high-level 0-to-1 design task into more specific facets, although at the cost of more cognitive load and time spent.
    \item Workflow: Compared to the depth-first way of ``getting \textit{a} design right'', users would find a breadth-first approach more appropriate for the 0-to-1 exploration although the large number of designs is challenging to process and keep track of.
    \item Output: While high fidelity generation is more realistic and polished, it also draws more attention to the low-level details (\eg colors and fonts); in contrast, low fidelity allows users to focus more on the high-level concepts and its lighter information load makes it more agile for further development.
\end{itemize}

\subsection{Participants}
\label{sec:participants}
We recruited participants in a large North American software company via three different internal communication channels used for recruiting participants interested in testing new GenAI products.
There were two inclusion criteria:
\one the role was either \uxd~ or \pm~ (whose work is most often related to 0-to-1 design \cite{10.1145/3715336.3735780}) and
\two the response to the screening question ``How is UX design related to your work?'' was either ``My main role is in UX design'' or ``UX design is a significant part of my work, though not my main role''.
There were no exclusion criteria.
In the end, we recruited
$24$ %
participants, $12$ \uxd s and $12$ \pm s, 
from $15$ different product teams.
Among them, there were
$13$ male, $8$ female, $2$ non-binary, and $1$ prefer-not-to-say; %
aged $30$ to $51$, %
with professional experiences ranging from $4$ to $20$ years.
We included both roles because \uxd s and \pm s are closely involved in early-stage, 0-to-1 design, albeit from different vantage points \cite{10.1007/11774129_15, MELEGATI2024107516}. Studying the differences between the two roles is beyond the scope of this work: professional role was not an independent variable, and the study was not designed to compare the roles. Our findings should therefore be interpreted as coming from two different but closely related perspectives on early-stage design, not as evidence of role-specific differences.
Participants were randomly assigned to one of four study sessions that took place between August 20 and 25, 2025.
Our study has been approved by the IRB of our institute and followed participants' company’s policy on engaging in research with human subjects.
Participants were paid a \$40 gift card for their time.

\subsection{Design, Tasks \& Procedure}
We employed a within-subject design. %
The within-subject factor was GenUI \textit{Tool}: Tool A (the version of \sysa \cite{stitch} used in the study) \vs Tool B (\sysb, our research prototype). These systems served as two usable, end-to-end instantiations of contrasting choices in input, workflow, and output fidelity. The comparison was not intended as a benchmark of either product or its underlying models; tool-level results describe the configurations participants encountered during the study.
A participant used each of these tools to conduct a different exploratory 0-to-1 design task and
{
was randomly assigned to one of the following study sessions where the order of the tools and the task-to-tool assignment was counterbalanced:
    \begin{enumerate} [leftmargin=1in]
        \item [Session \#1:] Task 1 (\sys) $\rightarrow$ Task 2 (\sysa)
        \item [Session \#2:] Task 1 (\sysa) \hspace{1.25em} $\rightarrow$ Task 2 (\sys)
        \item [Session \#3:] Task 2 (\sysa) \hspace{1.25em} $\rightarrow$ Task 1 (\sys)
        \item [Session \#4:] Task 2 (\sys) $\rightarrow$ Task 1 (\sysa)
    \end{enumerate}

We described each task to the participants in the following format (the two tasks only differed in the ``new application'' part):

\begin{quote}
    \it
    Your team is exploring a <new application>.

    Objective: Use the assigned GenUI tool to explore design ideas and propose one promising concept.

    Suggestion: start by fleshing out the task description before using the GenUI tool (you can use other tools).

\end{quote}

For Task 1's new application, we adopted and revised the one used in \cite{10.1145/3715336.3735780}---a mobile app that helps knowledge workers manage work-related stress.
Task 2's application was a desktop app that helps parents plan weekend activities with their children.
We intended for the two tasks to be complementary: Task 1 was a mobile app, focused on fostering personal health during day-to-day work; while Task 2 was a desktop app, focused on maintaining family connection on the weekends.

We started each study, after seeking informed consent from participants, with an introduction of GenUI and the purpose of this study, \ie comparing two GenUI designs.
Next, participants proceeded to the trials of using both Tool A and B.
Each tool started with a video tutorial demonstrating its basic interactive features and functionalities.
Next, we briefed participants with the task description, followed by information to access each tool.
Then, participants spent 20 minutes on the design exploration tasks using the tool.
We allowed them to use other tools if needed.
After each trial, participants fill out a questionnaire to rate their experiences on the input, workflow, and output of the tool.
Each session took place virtually on Google Meet and lasted about one hour.

Within two workdays after each design session, we conducted a 1:1 semi-structured interview with each participant to discuss their experiences and compare the three design differences between the two GenUI tools:
structured \vs unstructured input,
breadth- \vs depth-first workflow, and
low- \vs high-fidelity generation.

\subsection{Data Collection \& Analysis}

Immediately after each GenUI tool's design session, we collected participants' responses to 7-point Likert-scaled questions (Table~\ref{tb:survey}), corresponding to the three key aspects of comparison: input, workflow, and output.
We conducted Wilcoxon signed rank tests to understand whether there was any statistically significant difference in how they rated each GenUI tool.
Because comparing professional roles was beyond the scope of this work (\secref{participants}), all quantitative comparisons pool the 24 participants.

We collected qualitative data through the post-session interviews via taking notes of our discussions with participants.
We then employed a semi-structured thematic analysis \cite{terry2017thematic}.
The three key aspects formed a preliminary thematic structure to group the initial data points.
Within each thematic group, we then conducted further analyses, using \textit{benefits}, \textit{trade-offs}, and \textit{suggestions} as seed categories to guide a two-pass coding of participants' feedback.
The first author conducted the first pass of coding, which was then reviewed by one other author.
Disagreements between the authors were resolved through discussion, ensuring a comprehensive and consistent interpretation of the data.

\section{Findings: Usage Patterns of GenUI Tools}

To reveal usage patterns on the two GenUI tools, we analyzed screen recordings of participants' design sessions, which show how they exhibited different behaviors across various phases from initial input to mockup generation and to iteration (\tbref{usage-comparison}).
Further analyses of interaction logs on \sys examined its usage through a more detailed lens (\fgref{timelines}), showing how participants' time was distributed over different phases, how they balanced ``exploring the breadth'' and ``going deep'', and how they switched between these two activities.

\subsection{Findings from Screen Recording Analyses}
\label{sec:findings-recordings}

\begin{table*}[t]
\centering
\caption{\textbf{A comparison of typical usage patterns of the two GenUI tools}, based on a thematic analysis of screen recordings.}
\resizebox{\textwidth}{!}{%
\begin{tabular}[t]{p{0.12\textwidth} p{0.42\textwidth} p{0.46\textwidth}}
\toprule
&\sysa&\sysb\\
\midrule
\centering \textbf{Initial\\input}&
Heavy editing, going through the prompt back-and-forth multiple times &
Entering each input field once, with minimal or no edits\\
\centering $\downarrow$ &&\\
\centering \textbf{Pre-mockup-generation}&
Brief, single-step, reviewing an AI-generated list of screen descriptions& %
Elaborate, multi-step, selecting multiple options in the design space to generate and skim one or more specific designs\\
\centering $\downarrow$ &&\\
\centering \textbf{Post-mockup-generation}&
Surface-level browsing of mockups, occasionally delving into specific components on a screen&
Only viewing 1-2 designs' mockups, following the task flow to go through each screen\\
\centering $\downarrow$ &&\\
\centering \textbf{Mockup iteration}&
Frequent and low-level, across multiple places: follow-up prompts, customization panel, and comparing screens between iterations&
Few and mixed-level, either via adding to or modifying initial input or revisiting the design space to select different options\\
\bottomrule
\label{tb:usage-comparison}
\end{tabular}
}
\end{table*}%

As shown in \tbref{usage-comparison}, we compare participants' usage patterns of the two GenUI tools across four common stages of GenUI interaction: initial input, pre-mockup-generation, post-mockup-generation, and mockup iteration.

During initial input, one noticeable contrast is that \sysa users tended to edit their prompts heavily, going through the prompt back-and-forth multiple times.
For example, \uxd3 first briefly described an idea, then added stylistic descriptions (color schemes and layout);
\uxd9 edited the prompt twice---first adding layout details then navigation features.
In contrast, with \sysb's structured input, users entered each field once, with minimal or no edits.
Such differences suggest that \sysb users might have thought through their response to each field before entering it, whereas \sysa users might have treated the unstructured prompt as a working draft that resulted in the need to refine it iteratively.

Next, at the pre-mockup-generation stage, \sysa users performed a brief review of an AI-generated list of screen descriptions.
In contrast, \sysb users had to go through a more elaborate, multi-step process, going back-and-forth to select multiple options presented to them in the design space, before creating a specific design.
Some changed their design option selections multiple times and generated different designs, whereas others would move on to a design first and return later to the design space to explore other options (described below in the iteration phase).
For participants who created multiple designs at once, they would skim through all of them first (\pm3, \uxd2), although some paused mid-way to linger more on one design (\uxd8) or to open its UI view for more details (\uxd4).

After the tool generated UI mockups, both \sysa and \sysb users browsed through the screens at a surface level (\ie a quick scan).
The main difference is that \sysa users occasionally selected and zoomed into specific components on a screen whereas \sysb did not provide such ``deep-dive'' features.
\sysb users only viewed one or two selected designs' generated mockups, following the task flow to go through each screen. 

Then, in terms of iteration, \sysa users tended to make frequent and low-level changes across multiple places: follow-up prompts, customization panel (\eg adjusting fonts and colors), and comparing screens between iterations.
For example, \pm7 added layout requirements (grid) as a follow-up prompt and
\uxd1 adjusted both fonts and colors on the customization panel.
We did notice

the following participants attempted to explore high-level changes as new feature requirements, including advanced filtering function (\pm6), dashboard-style design (\uxd4), and map view (\uxd9). 

However, their attempts were unsuccessful as the tool either ignored them or failed to generate relevant results.
Participants seemed to have a tendency to be attached to the first design and
our analysis found no one started over with a new prompt to pursue a different design. 
In contrast, \sysb users made few and mixed-level changes mainly in two ways:
some went back to the input fields and added either stylistic (\uxd2, 4) or functional requirements (\uxd3, 9) and some changed their selections in the design space (\uxd2, 8, 9). 

\subsection{Findings from \sys's Interaction Log Analyses}
\label{sec:findings-logs}

To complement qualitative insights based on screen recordings, we further collected and analyzed interaction logs on \sys from all but two participants\footnote{Due to technical issues, they were unable to access and submit their browser's log file.}. 
By associating each logged event with a specific major component of the tool, we were able to break down the timelines of each participant's interaction with \sys (\fgref{timelines}).
Note that our logging mechanism did not employ a hard cut-off and, as a result, the lengths of participants' timelines varied: Some were much shorter likely because they started with using other tools (\eg Gemini) to prepare for the input, which we did not log; a few were much longer likely because they stayed in the tool or revisited it (\eg for answering survey questions) even after the session was concluded.

\begin{figure*}[t]
\centering
\includegraphics[width=1\textwidth]{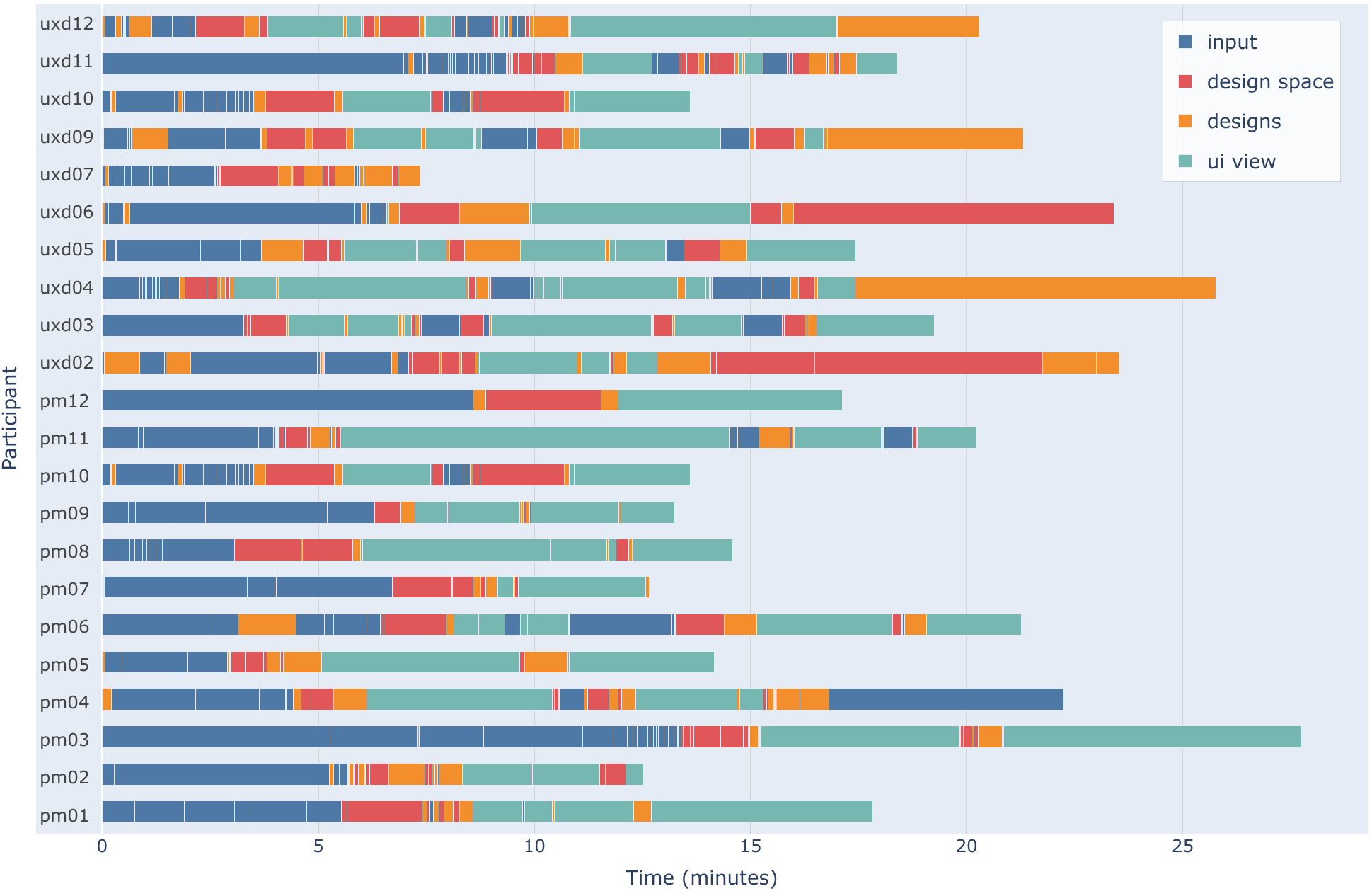}
\caption{
\textbf{A time-series visualization of participants' interaction with \sys}.
Participants spent various amounts of time on the four different parts/steps of \sys over the course of their design session ($\mu=18.1~\text{min}$): a significant amount of time on the structured input ($\mu=5.9~\text{min}$), followed by switching back and forth between exploring the breadth (design space, designs: $\mu=5.5~\text{min}$) and going deep (UI view: $\mu=6.7~\text{min}$). 
}
\label{fg:timelines}
\end{figure*}

Many participants spent quite a portion of their time entering the structured input.
Further analyses revealed that participants spent an average of $5.9$ minutes ($\text{SD}=2.9$ minutes) on input, which is about one third of their entire session (average is $18.1$ minutes).
Noticeably, some participants (\eg \pm7, \pm12, and \uxd11) spent over half of their sessions on input.
Such behaviors suggest that entering structured input is a non-trivial, or even time-consuming step, which we discussed later in the interview findings.

The rest of participants' timelines were divided between interacting with three components: the design space panel, the designs (presented as design cards), and each design's UI view.
By and large, we can consider interacting with design space and designs as ``exploring the breadth'' whereas delving into a design's UI view as ``going deep''.
Further analyses show that participants spent an average of $5.5$ minutes ($\text{SD}=3.5$) exploring the breadth and an average of $6.7$ minutes ($\text{SD}=3.1$) going deep.
Some participants were more breadth-focused, spending over $2\times$ the amount of time on design space and designs as on UI view (\uxd2, 6, 7) whereas quite a few were more depth-focused, spending over $2\times$ the amount of time on UI view as on design space and designs (\pm1, 3, 8, 9, 11; \uxd3).
These patterns, also visible in the role-ordered timelines in \fgref{timelines}, are descriptive only: the role subgroups were too small to test for differences, so we do not interpret them as role effects (\secref{participants}).
To understand how participants switched between breadth and depth activities, we counted the segments of UI view interaction.
Intuitively, having multiple UI view segments means the user frequently goes deep and then switches back to breadth.
On average, participants had 3.0 UI view segments ($\text{SD}=1.4$), \ie switching between breadth and depth three times.

\begin{table*}
    \centering
    \caption{\textbf{Results of the post-session survey.} 
        We asked each participant to rate (on a 1 to 7 Likert-scale) their agreement or disagreement with each statement regarding both GenUI designs.
        The heatmap visualizes the distributions of their ratings.
        We then conducted a Wilcoxon signed-rank test on participants' ratings between the two GenUI designs. Ratings pool all 24 participants. }
        \vspace{-1em}
    \includegraphics[width=\textwidth]{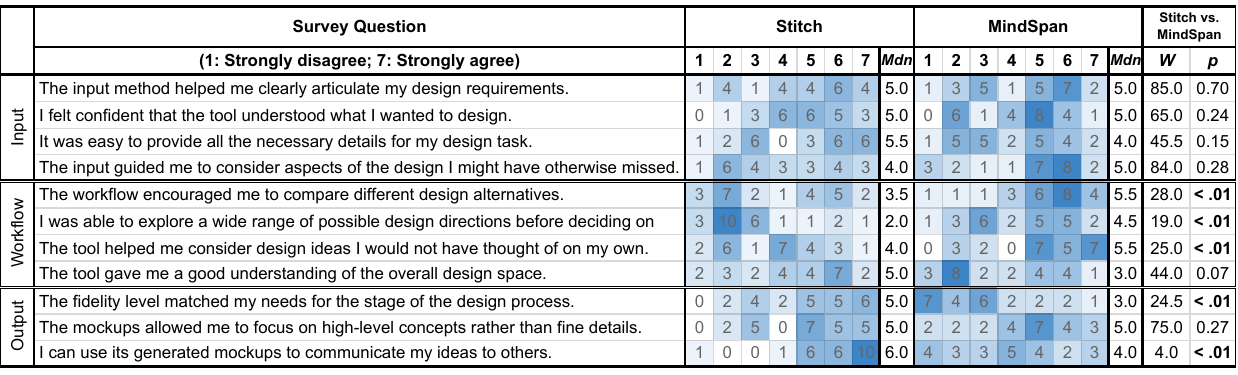}
    \label{tb:survey}
\end{table*}

\section{Findings}
\label{sec:findings-pros-cons}

For each of the three design aspects, we first summarize quantitative results of survey questions and then discuss recurring themes distilled from qualitative interview data.
As explained in \secref{participants}, our findings come from two different but closely related perspectives on early-stage product and UX design. The \pm~ and \uxd~ labels preserve sample context and identify the source of quotations; they do not denote analytic groups. Accordingly, we report findings across both roles and do not draw conclusions about differences between them.
Likewise, \sysa and \sysb identify the two configurations evaluated in the study rather than stable product capabilities. Because commercial services and underlying models can change, our claims concern participants' experiences with those configurations and the design dimensions they instantiated; they are not rankings of current products or models.

\subsection{Unstructured \vs Structured Input}
\label{sec:findings-input}
We asked four input-related questions during the post-session survey (\tbref{survey}).
Statistical analyses show that, for each of the questions, there is no significant difference between participants' ratings of the two GenUI input designs.
As analyzed below with qualitative data, the lack of difference in ratings likely reflects that each way of input has its own strengths and struggles for expressing UX design needs, even though most participants preferred to start exploration with unstructured input.

Participants
appreciated how the structure surfaced key UX-specific facets and \inquo{clearly lays out the things that will control the output}{\pm11}, making them think through the problem harder (\uxd3-5, 9; \pm6, 11).
The structure is more than a template (\uxd5, \pm3); its specificity served as a scaffold for participants to deepen their thoughts:

\quo{Initially did not differentiate tasks from goals; having this structure forces me to think at a more granular level.}{\uxd9}

\quo{Tasks helped to bridge the design to flows and allow me to focus on one part of the experience and jump to others—serving as a structure to keep me aware of which parts of the user journey are in the design.}{\uxd8}

Ten participants described difficulty or a lack of readiness when starting exploration with structured input, particularly when filling out ``tasks''.
Foremost, from a UI point of view, participants felt the ``form filling'' experience too heavy (\pm6), overwhelming (\pm2), or even intimidating (\uxd1),
Participants felt such a structure \inquo{harder to get started (with)}{\pm8} for design exploration---
\inquo{I haven't soaked in the idea; do not want to think about these things before visualizing something}{\pm1}.
The fundamental problem is that, to some participants, asking for specific information upfront felt premature and lacking openness, \inquo{forcing}{\pm3} them to think through many aspects before they were ready (\pm1, 3, 6).
Further, some participants also had initial ideas that \inquo{don't fit any of those boxes}{\uxd5} and they did not know where to input those kinds of information (\pm7).

\textbf{In contrast, 19 participants found unstructured input, though imperfect, more appealing because of a more open format and a lower barrier of entry}.
Participants did point out many challenges of unstructured prompting that echoed findings from prior work \cite{10.1145/3613904.3642016, 10.1145/3544548.3581388, chen2023next, 10.1145/3613904.3642754, 10855599}, \eg difficulty of formulating effective prompts (\pm2, 11; \uxd2, 5, 6, 8, 11, 12) and the uncertainty of not knowing what or how much to specify in the prompt (\pm3; \uxd1, 3, 6).
Nonetheless, they preferred this way of input upfront for its openness (\pm2, 10; \uxd4, 10, 11) and flexibility (\pm4, 7, 10; \uxd4, 9) that could help them get started easily (\pm1, 8, 11).

\subsection{Depth- \vs Breadth-First Workflow}
We asked four questions on how each GenUI tool's workflow supported the assigned design exploration task,
based on the rationale that diverging to alternatives is key to prevent fixation at the early stage \cite{designcouncil_framework_innovation, 10.1145/1978942.1979359, goldschmidt2016linkographic}.

As shown in \tbref{survey}, participants rated \sysb higher on the following statements:
\one ``\textit{The workflow encouraged me to compare different design alternatives}'' ($W=28.0$, $p<.01$);
\two ``\textit{I was able to explore a wide range of possible design directions before deciding on one}'' ($W=19.0$, $p<.01$), and;
\three ``\textit{The tool helped me consider design ideas I would not have thought of on my own}'' ($W=25.0$, $p<.01$).
These ratings indicate that participants did benefit from using the breadth-first \sysb for design exploration.
Interestingly, we did not find a statistically significant difference in the fourth question ``\textit{The tool gave me a good understanding of the overall design space}''.

Overall, participants recognized the benefits of the breadth-first workflow (\uxd6, 10, 12; \pm2, 6, 8, 10, 12).
Dissecting the design task into different dimensions in the design space allowed them to \inquo{think about combinations of things}{\uxd4} and to \inquo{mix and match different ways of responding to design needs}{\uxd9}.
Such an early exposure to a \inquo{wider field}{\pm10} is important for the following reasons.

UX design is highly multi-faceted and an exploration often requires considerations of users' persona, product metaphor, mental models of task flow, among many other things.
\textbf{Ten participants described how a breadth-first workflow challenged their (implicit) assumptions on these UX design facets}, pushing them to 
research (\pm2) and test hypotheses (\pm3), seeing the \inquo{polarity}{\uxd4} of ideas.
As a result, the breadth-first workflow allowed participants to explore ideas they would not have thought of (\pm9, 12; \uxd3, 5), \eg
\inquo{different factors that made experiences different from one another}{\uxd2}, as well as
user journeys \inquo{different than what I would've thought of}{\pm8} and that they \inquo{didn't think through when getting started}{\uxd8}.

\quo{.. the (\sys) app more full ...
[It] gave you these spectrums [of] target directions, helping you think through the breadth of what the app could be, felt more complete.}{\uxd4}

\quo{... the tool made me think about the aspect of AI's persona--would not have come to if I had used the other tool.}{\pm3}

\textbf{Four participants also treated a breadth-first workflow as a thoughtful, analytical, and unhurried engagement with the design problem}. It allowed them to take a slower (\uxd1, 3) pace, \inquo{go through a more granular process}{\pm8} where the questions in the design space helped them \inquo{step back and look at the big picture}{\uxd3}.

\quo{... slower pace [with] different directions of problem solving: stop, pause---we got the basic here; these are the things I will think about before interviewing users.}{\uxd3}

\quo{Like a sounding board: instead of saying ``what you have is perfect, let's design it,'' it asks ``what if this'' and ``... that''.}{\pm12}

In terms of trade-offs, participants noted how the breadth-first workflow took longer (\uxd11), more steps (\uxd1), and a lot of \inquo{pogo stick like}{\uxd9} back-and-forth (\pm7) before they got to see some UI mock-ups.
To navigate the design space, participants mentioned the sheer number of options, having to manage the possibilities of 3 questions $\times$ 4 options each (\pm1, 5; \uxd10).
The sticky-note-sized, bullet-pointed text descriptions of each design idea,
felt \inquo{too busy}{\uxd10} and \inquo{too technical ... text-heavy}{\pm2, 6} and the freeform placement lacked organization (\pm6), hard to navigate (\pm6, \uxd5). 
A lack of visuals (\pm2, \uxd11) created a barrier for design exploration:
\uxd6 mentioned how they were used to seeing design as mockups rather than the textual design cards and \uxd11 felt an exploration was inadequate without actually generating the mockups.

\textbf{Indeed, 11 participants described the challenge of effectively and efficiently previewing UX design ideas at a glance}.
While some other generated contents can be browsed at a glance of their snapshots (\eg sentences \cite{10.1145/3613904.3641899}, images \cite{conf/uist/EvirgenC22} and 3D models \cite{10.1145/3173574.3173943}), it is unclear how to present a single preview of a 0-to-1 UX design. 
Each UX design often consists of multiple screens to support multiple task flows: presenting all of them in one visual storyboard is too overloaded (and time-consuming to generate); only showing one screen is feasible but cannot fully represent the entire design.
We discuss later in \secref{implications-breadth} possible solutions to address this challenge.

In comparison, \textbf{five participants felt the depth-first workflow excels at speed and efficiency}. 
Participants liked how it produced a design fast (\pm7, \uxd1), which felt like making quick progress (\pm5, 10), getting \inquo{closer to done faster}{\pm10}.
This efficiency-oriented preference prioritized making progress on fewer design options over researching a broader set of alternatives (\pm2, 5, 10).
\textbf{However, 13 participants pointed out that the depth-first workflow remains limited in supporting design exploration}---\inquo{not as blue sky and wildly creative as breadth-first approach}{\pm10} and \inquo{didn't really help me explore}{\pm9}.
Foremost, delving into a design right away was
\inquo{too focused on one solution}{\uxd11} while it remains
\inquo{unclear what to do next with the generation}{\uxd3}.

\quo{It talked back to me what I was thinking and gave me one option, asking if I wanted to change anything, as opposed to suggesting other options. I am still limited to what I was thinking about, not outside of the box.}{\pm12}

Further, participants described a tendency to focus on \inquo{look and feel}{\uxd4} and \inquo{tactical details}{\uxd3}, \eg colors and fonts, when given the one design generated upfront.
The depth-first approach seemed to have \inquo{framed my mindset into focusing on visual refinement}{\uxd4} and \inquo{made me think we could come up with something 'done' and just finetune it}{\uxd3} even though they realized they should \inquo{instead focus on flow and whether the right information is placed}{\uxd10}.
Such feedback was also related to the fidelity aspect, which we discuss below.

\subsection{High- \vs Low-Fidelity UI Generation}
\label{sec:findings-fidelity}
As shown in \tbref{survey}, 
participants rated the high-fidelity approach higher on the following two questions:
\one ``\textit{The fidelity level matched my needs for the stage of the design process}'' ($W=24.5$, $p<.01$); and
\two ``\textit{I can use its generated mockups to communicate my ideas to others.}'' ($W=4.0$, $p<.01$).
As discussed below, such preferences were largely framed by participants' professional practices.
Participants gave a median rating of $5$ for both tools (no significant difference) for the question ``\textit{The mockups allowed me to focus on high-level concepts rather than fine details}''. 
To many participants, low fidelity's lack of fine details was also an impediment in conveying the design concepts.

Echoing prior work \cite{doi:10.1177/154193128903300405,10.1145/223904.223910,buxton2010sketching, 10.1145/1357054.1357074},
nine participants did recognize many positive aspects of low fidelity, \eg its appropriateness for 0-to-1 design exploration (\pm11; \uxd8, 10), its focus on interactions (\pm3, \uxd12) while deferring judgments on details (\uxd2, 10), and the freedom and flexibility knowing that it is not a finished product (\pm9). 
Participants also showed a clear understanding of high fidelity's drawbacks, \eg its details become distractions from the core design (\pm3, 4), drawing too much of people's attention to minutiae (\pm9; \uxd3, 10).

\textbf{Despite the many reasons to favor low fidelity for design exploration, ten participants still mentioned a preference for high fidelity}.
Besides the obvious reason that high fidelity being more visually and interactively appealing (\pm4, 10, 11; \uxd9), participants also mentioned their desire for getting to a high-fidelity end-state fast (\pm3, 5, 6, 11; \uxd5, 10) where advanced AI can help them skip the previous need to convert from low to high fidelity (\pm2, 11; \uxd5).
Such mindsets echoed the earlier finding of how some participants preferred the depth-first workflow because of its speed and efficiency. 

\textbf{Fourteen participants indicated that high fidelity better fits their professional UX practices whereas low fidelity was considered unwieldy}.
First, participants considered high fidelity more compatible with UX work, providing a \inquo{good version to start working off of}{\pm11} and better for hand-off to downstream tasks (\uxd4). 
Contrary to our hypothesis that low fidelity's light information load makes its further development more agile, participants considered low fidelity \inquo{slows down the design process}{\pm3} and \inquo{harder to build off (of)}{\uxd10}.
Such judgments seemed to have stemmed from participants' tendency to measure the distance between the low-fidelity and the eventual state, leading to the mindset that higher fidelity equals more progress and closer to the finish line.

Second, working with high-fidelity mockups felt familiar to the professional tools participants used, \eg Figma (\uxd2, 5, 6), and closer to how they \inquo{use design systems in work}{\uxd8}; in contrast, low fidelity felt outdated to them, \eg like early wireframing tools (\pm6, \uxd10), and \inquo{unlike real design systems}{\uxd5}.

Finally, high fidelity serves better for other stakeholders, \eg \inquo{for engineers and power users to look at detailed and more engaging mock(ups)}{\pm3}, for showing a live demo (\uxd11), and for getting feedback from others (\uxd2). In contrast, low fidelity is \inquo{hard to explain to a user}{\uxd1} and cannot be handed off due to its openness that causes ambiguity (\pm5), thus requiring more UX work and resources (\pm5, 8).

\textbf{Such a ``high-fidelity mindset'', combined with a high expectation built up through the breadth-first exploration, amplified participants' disappointment when presented with low-fidelity mockups}.
Having developed a clear understanding of user journey and tasks (\uxd8) and explored various designs in textual forms (\pm12), participants felt seeing a low-fidelity output mismatched the ideas that had emerged in their mind (\pm12)---a \inquo{letdown}{\pm6} that \inquo{made it hard to translate those (design ideas) to visuals}{\uxd8}, \inquo{undoing the effort spent on understanding the user and flow}{\uxd3}. As one participant put it--- 

\quo{Fidelity too low; uninspiring compared to the quality of thoughts leading to it—cannot match the promise of the design ideas.}{\uxd9}

\section{Implications for Design}

In this section, we distinguish between findings observed in our study and design implications derived from them. The implications are not directly validated design solutions; rather, they are cautiously framed directions suggested by participants' experiences and by our interpretation of the trade-offs surfaced in the study. They concern interaction choices rather than assumptions about the capabilities of future models.

\textbf{Finding: participants preferred an unstructured start, yet valued the guidance of structured input}.
Survey ratings of the two input designs did not differ (\secref{findings-input}), yet most participants (19 of 24) preferred to start exploration with unstructured input, while several also credited structured input with helping them think through key facets of UX design.
Participants' suggestions often gravitated towards {a hybrid, mixed-initiative \cite{horvitz1999principles}, unstructured-to-structure approach}, but our study did not directly evaluate such a hybrid design.
Such a hybrid input dialog may be especially useful for participants who are not yet ready to specify user goals and tasks at the outset of exploration.
\textbf{Implication: GenUI may benefit from mechanisms that let users begin openly and add structure gradually}.
As in conversational AI interfaces that augment a vanilla prompt with suggestive chips or autocompletion, GenUI
could \inquo{suggest a more structured way to enter the open-ended prompt}{\uxd5}.
As users start typing a prompt, the tool might proactively map contents to specific fields (\uxd1), or reactively let
the user select an option to rephrase or refine the prompt with more structure (\pm5, \uxd5), where the tool could in turn suggest missing information (\pm12).
Alternatively, users can further defer formalizing the critical user journey, perhaps explore some designs first, and later build on their initial unstructured input with \inquo{a guided conversation about users, goals, and tasks}{\uxd12}.
For users who prefer a ``warm start'', these mechanisms may make it easier to react to something tangible before developing a structural understanding upfront.

Two systems illustrate how such incremental structure might be built. Park \etal let users attach and inspect design semantics on top of a generated UI, so structure accrues around an artifact rather than preceding it \cite{park2026bridging}; Xu \etal reify intents as separately editable artifacts, so a user can formalize one facet at a time instead of committing to a complete specification upfront \cite{xu2026ooprompt}.
Both suggest that the barrier we observed stems less from structure itself than from requiring all of it before anything is generated.

\textbf{Finding: breadth-first GenUI made previewing many UX designs a central challenge}.
\label{sec:implications-breadth}
Our findings suggest that the crux of breadth-first GenUI may be effectively and efficiently previewing UX designs, each of which itself often consists of multiple screens to support multiple tasks.

This compounds a limitation already noted for single-design GenUI tools, namely their focus on individual static screens at the expense of experiences spanning multiple screens \cite{lu2024flowy}; going breadth-first multiplies that problem by the number of alternatives.
It also raises a question of whether breadth is real: prior work cautions that repeatedly sampling an LM tends to return candidates from the same semantic neighborhood, so a system must actively diversify rather than merely generate more \cite{ibrahim2026anchorless}.

\textbf{Implication: future GenUI systems could explore richer ways to summarize and compare alternatives}.
Three possible directions emerged in the discussions with participants:
\one 
{Start depth-first, generate a ``seed design'' as an anchor, and then extrapolate the breadth of other possibilities}---\inquo{give me something tangible, then try different variations to suggest ideas I hadn't thought of---classic brainstorming}{\uxd3}.
\two 
Show design alternatives side-by-side via a competitive analysis \cite{rosson2002usability} table:
To realize this approach, a system would likely need to identify the critical components or interaction flow on each UI that are worth comparing, as well as highlight key differences, \eg using a mix of modalities---textual summarization and visual annotations on the UIs.
\three Generate a rapid serial visual presentation (RSVP) \cite{potter2018rapid} of UI states---a brief animation of the interaction flow as the preview of a design idea.
Ideally, in just a few seconds, a user could watch a few key frames and grasp how a task is performed using the generated screen mockups.
Rather than rendering each screen in full detail, the RSVP could adopt an abbreviated version, \eg only showing the layout and using icons to signify \cite{norman2008signifiers} each component's functionality.

\textbf{Finding: participants often expected generated mockups to look closer to product-ready artifacts}.
In the study context, participants appeared to treat polished, near-product-like mockups as the norm for AI-generated artifacts. Their ``high-fidelity mindset'' may have been shaped by exposure to product-like AI prototyping, including ``vibe coding'' \cite{sarkar2025vibe}.

This expectation is not idiosyncratic to our participants. Kobiella \etal observed the same shift across technical skill levels, where prototypes previously treated as disposable were expected to be carried forward \cite{kobiella2026throwaway}, and Li \etal found product teams reorganizing around the speed of AI-generated artifacts at the cost of reflection \cite{li2026vibeteams}.
Read against Hundhausen's argument that classic fidelity wisdom rested on high fidelity being expensive \cite{hundhausen2026fidelity}, our finding suggests that the rationale for low fidelity need not rest on cost or effort; it can instead be articulated in terms of what a designer gains by withholding detail.

Critiques of design practice offer a structural reading of this mindset.
Kolko described interaction design becoming a commodity, operationalized as the assembly of safe, predefined components rather than open-ended ideation \cite{10.1145/3736591}, and Visci\`{o}la argued that human-centered design has been subordinated to business imperatives that prioritize rapid output, constraining designers' room for speculative exploration \cite{10.1145/3733153}.
Such production-oriented pressures may partly explain why participants rated high fidelity more favorably, although our study did not measure participants' organizational orientations.
Conversely, the value that some participants saw in low fidelity resonates with post-growth HCI, which calls for research, design, and practice that prioritize quality of life over growth \cite{10.1145/3624981}.

\textbf{Implication: GenUI systems should make fidelity adjustable and purpose-linked rather than encode either level as a permanent default}.
Rather than treating fidelity as a binary choice or prescribing a fixed starting point, systems could support mixed fidelity---``high-fidelity on some dimensions and low-fidelity on others'' \cite{10.1145/1124772.1124959}---and let users align the initial level with their design purpose. A high-fidelity starting point may serve users who need a near-end-product view, whereas lower fidelity may serve users who want to defer detail during exploration.
Meanwhile, enabling mixed-fidelity would allow users to dial up or lower the overall fidelity based on their preference, as demonstrated in early work on a mobile UI design tool \cite{10.1145/1385569.1385606}. 
Further, rather than considering fidelity as a global parameter, future systems could let users select a subset of the UI components to locally change their fidelity.
For example, low fidelity might be enough for conventional UIs (\eg buttons and text boxes), but users might require more details to better understand the novel components, \eg a treasure map for parents and children to discover weekend activities.
Future developments of GenUI tools could investigate how to decide which components should be rendered at what level of fidelity.

\textbf{Implications for other AI-powered creativity support tools}.
First, many other design tasks are also defined by multiple facets (\eg form, function, and cost for furniture design). Thus a hybrid, mixed-initiative, unstructured-to-structured input may help users ease into exploration while staying aware of what important information they might consider.
Second, like UX, various other designs cannot be easily previewed at a glance, \eg narrative media (films, games, novels), interior designs, and services.
To address this, future systems could explore creating one design first and then extrapolating, comparing alternatives side-by-side, or using rapid serial visual presentations.
Third, fidelity is a general notion in other design domains, \eg the spectrum of sketches, scaled models, and photorealistic rendering in architecture.
This suggests that mixed-fidelity generation may be a useful technique for allowing professionals to adjust the kinds and levels of detail they focus on based on the stage of their design process.

\section{Limitations, Future Work, \& Conclusion}

Our study was intentionally scoped to early-stage, individual design exploration. We recruited UX designers and product managers because both are closely involved in early-stage design, albeit from different perspectives. Studying how the two roles differ was beyond the scope of this work; our findings should be read as coming from these two different but closely related perspectives rather than as evidence of role-specific differences, which future work could examine with a sample designed for that comparison. Descriptively, some themes did lean toward one role: most participants who found structured input difficult to start with were product managers (8 of 10), the efficiency-oriented preference for depth-first generation came from product managers, and the tendency to focus on visual details under depth-first generation came mainly from UX designers; such leanings could help motivate that comparison. Recruiting participants from a single large company following prior work~\cite{subramonyam2024content, 10.1145/3715336.3735780} bounds the generalizability of our findings. Participants came from 15 diverse teams, yet their shared organizational culture may have shaped their expectations of GenUI. The sample also does not represent other professionals who participate in UX work. Future work should therefore expand across companies and industries, as well as to other UX-related roles such as front-end \swe, \uxr, and cross-functional collaborators.

Our experimental setup also emphasized exploration over the broader arc of UX work. We scoped tasks to 0-to-1 design contexts where GenUI has shown strong potential~\cite{10.1145/3715336.3735780}, leaving open how our findings extend to more common n-to-(n+1) iteration. Similarly, our 20-minute exploration sessions enabled participants to try two tasks within an hour, but may not reflect longer-term use in which ideas are revisited, refined, or abandoned over days and weeks. Because the survey questions on depth- \vs breadth-first workflows focused on divergent exploration, they did not assess whether GenUI also helps users develop ideas more fully or in greater depth. Relatedly, we examined exploration processes and trade-offs rather than evaluating the quality of the final design outcomes produced. Longitudinal studies that track both process and outcome quality would complement our findings.

Finally, our comparison surfaced trade-offs between two coherent GenUI paradigms rather than isolating each design factor independently. Bundling input structure, exploration workflow, and output fidelity made the tools meaningful to use as end-to-end systems, but also made it difficult to attribute observed effects to any single dimension. The tools also differed in maturity: one was a commercial beta product, while the other was our research prototype. In particular, our SVG-based implementation of low-fidelity generation may have influenced factors beyond fidelity itself, such as perceived system capability or instruction-following quality. Future studies could compare tools with more similar levels of polish, or use controlled prototypes that vary one design dimension at a time. Moreover, because our study focused on individual exploration, it does not capture the collaborative nature of UX work; future work should examine how GenUI supports group ideation, coordination, and co-creation.

The study also captures a time-bounded technological configuration. Sessions took place in August 2025 using the then-available Stitch beta and the Gemini 2.5 model family documented in \secref{implementation} and the Appendix. Commercial services and models can change after deployment; therefore, the findings should not be read as a current product benchmark or as a claim that the same tool-level ratings will reproduce with later versions. We use these configurations as probes of recurring design choices. Replications with newer models and controlled interfaces could distinguish effects mediated by input, workflow, and fidelity from effects mediated by generation capability.

GenUI can inherit a chat-based interaction pattern: start from an unstructured prompt, delve into one design, and generate a polished mockup. We made that pattern visible as a set of design choices by contrasting unstructured \vs structured input, depth- \vs breadth-first workflow, and high- \vs low-fidelity generation. This comparison surfaced durable tensions: openness \vs guidance when specifying intent, efficiency \vs reflection when traversing alternatives, and polish \vs abstraction when representing an idea. The observed resistance to up-front structure, difficulty of previewing multi-screen alternatives, and preference for high fidelity are time-bounded empirical findings; the contribution is the design-level account that relates those observations to choices future systems must still make.
By treating named systems and models as apparatus rather than the contribution, this work offers a vocabulary for comparing and calibrating GenUI interactions as implementations evolve, with relevance to AI-powered creativity tools beyond UI design.

\begin{acks}
Our colleagues Ed Hurst, Yuling Gao, Michael Xieyang Liu, and Michael Terry provided invaluable feedback and support to this project.
We thank our collaborators and study participants for their contribution to this research.
We are also grateful to reviewers who provided valuable feedback to help us improve this work.
\end{acks}

\bibliographystyle{ACM-Reference-Format}
\bibliography{main,xac}

\appendix

\section{Step-by-Step Implementation Details of \sys}
\label{sec:implementation_details}

\subsection{Generating Design Space}

\noindent
\textbf{Step 1}: Brainstorm multiple ideas \\
\textbf{Model}: \textsf{gemini-2.5-flash-lite} \\
\textbf{Prompt}:
{\footnotesize
\begin{Verbatim}[breaklines,fontsize=\scriptsize]
SYSTEM:
You are an expert UX strategist and creative thinker. 
Generate fundamentally different UI design ideas for the given context.

ASSISTANT:
Brainstorm at least 15 distinct product metaphors or paradigms 
to inform the design of this application. Each idea should represent 
a different conceptual approach to solving the user's problem.

Inputs:
- Context: {context}
- User: {user}
- Goal: {goal}
- Tasks: {tasks}
- Examples: {examples}

Requirements:
- At least 15 ideas, relevant to the context, user, goal, and tasks
- Draw inspiration from provided examples and user comments,
  but do not be constrained by them
- Focus on high-level conceptual approaches, not implementation details

Output Format:
Return a JSON array where each idea object includes:
- idea_id: Unique integer identifier
- idea_name: Short, descriptive name
- description: Brief explanation of how this idea would work
- inspiration: Relation to the context, user, goal, or examples

Constraints:
- Response must be a valid JSON array
- Do not include any text, code fences, or annotations before/after the JSON
- Each idea must be conceptually distinct
\end{Verbatim}
}

\vspace{1em}

\noindent
\textbf{Step 2}: Identify design dimensions\\
\textbf{Model}: \textsf{gemini-2.5-flash-lite} \\
\textbf{Prompt}:
{\footnotesize
\begin{Verbatim}[breaklines,fontsize=\scriptsize]
SYSTEM:
You are an expert UX strategist. Create a high-level conceptual
design space from the provided divergent ideas, using descriptive,
plain language understandable to non-designers (e.g., product
managers, software engineers, UX researchers).

ASSISTANT:
Follow these steps:

STEP 1 – Analyze Ideas
- Review all ideas to identify themes, tensions, and patterns
- Highlight fundamental UX-related differences

STEP 2 – Identify Dimensions
- Derive 3 orthogonal design dimensions, each reflecting a key tension,
  role, or workflow shift
- Each dimension should express a meaningful choice that shapes the user
  experience
- Write one sentence explaining why each dimension matters

STEP 3 – Create Options
- For each dimension, define 3–5 clearly distinct options
- Each option should represent a different approach along the spectrum
- Provide one-sentence plain-language descriptions

Inputs:
- Divergent Ideas: {divergentIdeas}
- Context: {context}
- User: {user}
- Goal: {goal}
- Tasks: {tasks}
- Examples: {examples}

Requirements:
- Exactly 3 design dimensions, each distinct and relevant to the context
- Each dimension’s description should be concise, ideally a question that
  prompts reflection
- Options must use plain, accessible language and represent meaningful
  alternatives
- Exclude low-level design details (e.g., styling, information density)
- Use divergent ideas, examples, and user comments as inspiration

Constraints:
- Response must be a valid JSON array
- Do not include text, code fences, or annotations before/after the JSON
- Each dimension and its options must be clearly distinct and relevant
\end{Verbatim}

}

\noindent The reference to non-designers in this prompt asks the model to describe the design space in plain language, lowering the barrier to engaging with it; it does not restrict \sys to non-designer users.

\subsection{Generating Specific Designs}

\noindent
\textbf{Step 1}: Generate design concepts\\
\textbf{Model}: \textsf{gemini-2.5-flash-lite} \\
\textbf{Prompt}:
{\footnotesize
\begin{Verbatim}[breaklines,fontsize=\scriptsize]
Analyze the provided information and generate 5 distinct high-level
design concepts that help the user achieve their goal by performing
the tasks. Each concept must directly address the following:

Design Parameters: {designParameters}

Output Format:
Return a JSON array where each design object includes:
- design_id: Unique integer identifier
- design_name: A highly descriptive short sentence that conveys the
  design without needing further explanation
- core_concept: Main design approach, showing how it addresses each
  parameter (use bullet points)
- detailed_description: A detailed explanation of the concept for a
  UX designer to implement

Constraints:
- Response must be a valid JSON array
- Do not include text, code fences, or annotations before/after the JSON
- Incorporate user comments to align with preferences and feedback
\end{Verbatim}

}

\vspace{1em}

\noindent
\textbf{Step 2}: Generate task-wise screen descriptions\\
\textbf{Model}: \textsf{gemini-2.5-flash-lite} \\
\textbf{Prompt}:
{\footnotesize
\begin{Verbatim}[breaklines,fontsize=\scriptsize]
Analyze the overall design of the application. For each task,
generate a 2–3 sentence description of the screens needed to complete it.
Each description must specify the screen’s purpose, UI elements, and functionality.

Inputs:
- Overall Design: {overallDesign}
- Tasks: {tasks}

Each screen description should include:
- Purpose: What the screen is for and what users accomplish
- Elements: Key UI components (buttons, forms, lists, navigation, etc.;
  can be nested objects or arrays)
- Functionality: Actions users can perform and how the screen behaves
- Layout: Organization and positioning of elements
- Interactions: How users navigate to/from this screen and interact

Descriptions must be detailed enough for a UX designer to create the screen.
\end{Verbatim}

}

\vspace{1em}

\noindent
\textbf{Step 3}: Generate unified screen descriptions\\
\textbf{Model}: \textsf{gemini-2.5-flash-lite} \\
\textbf{Prompt}:
{\footnotesize
\begin{Verbatim}[breaklines,fontsize=\scriptsize]
You are a UX design assistant helping define the core structure of an application. 
I will provide several user tasks, each with a sequence of screens. Analyze these 
task flows and propose a unified set of conceptual screens that support all tasks, 
while minimizing redundancy and preserving key functionality. Use as few screens as possible.

Output high-level screen concepts, not implementation details. Focus on each screen’s 
role in the workflow, core elements, and essential interactions. Avoid detailed layouts 
or exhaustive UI specifications.

For each screen, include:
- title: Short, descriptive name
- purpose: What users achieve and why it is necessary
- core_elements: Major components essential to the screen’s purpose (e.g., search bar, 
  content list, form section)
- key_interactions: Main actions users can perform and navigation to/from this screen
- data_notes: (Optional) Data shown or collected, only if essential

Task-Specific Screen Descriptions:
{screenDescriptions}

Requirements:
- Return a JSON array of screen objects that unify overlapping functionality, clarify 
  intent, and remain general enough for low-fi prototyping
- Each screen object must include: title, purpose, core_elements, key_interactions, 
  and optional data_notes
- Response must be a valid JSON array with no text, code fences, or annotations before/after
\end{Verbatim}

}

\subsection{Generating UI Code and Taskflows}

\noindent
\textbf{Step 1}: Map screens to each task\\
\textbf{Model}: \textsf{gemini-2.5-flash-lite} \\
\textbf{Prompt}:
{\footnotesize
\begin{Verbatim}[breaklines,fontsize=\scriptsize]
Given the following tasks and screen descriptions, determine which
sequence of screens is needed to complete each task. Use 0-based
indices to reference screens.

Tasks:
{tasks}

Screen Descriptions:
{screenDescriptions}

For each task:
- Identify the required screens in order
- Include indices of all necessary screens; each task must use at least one
- Indices cannot exceed the number of available screens
- Typically, a screen does not repeat within the same task
- Each screen must be used by at least one task

Output Format:
Return a JSON object with a "tasksWithScreens" array. Each entry includes:
- task: the task description
- screens: ordered array of screen objects
  - screen_id: index of the screen
  - interaction: brief description of what the user does on this screen
                 to progress to the next

Constraints:
- Response must be a valid JSON object
- Do not include any text, code fences, or annotations before/after the JSON
\end{Verbatim}

}

\vspace{1em}

\noindent
\textbf{Step 2}: Generate UI code for each screen\\
\textbf{Model}: \textsf{gemini-2.5-flash-lite} or \textsf{gemini-2.5-pro} \\
\textbf{Prompt}:
{\footnotesize
\begin{Verbatim}[breaklines,fontsize=\scriptsize]
ROLE & GOAL:
You are an expert UI wireframing assistant. Produce one complete,
valid SVG that looks like a hand-drawn wireframe of the screen
described below.

SCREEN DESCRIPTION:
{screenDescription}

REQUIREMENTS:
- Include every UI element listed in "elements"
- Implement all behaviors in "functionality"
- Reflect user comments and preferences in layout and style

STYLE & FORMAT RULES:
1. Root <svg> must include a proper viewBox
2. All attribute values must be fully quoted (d, x, y, width, height, etc.)
3. No unclosed quotes or stray characters
4. Each <path> must contain a complete d="..."
5. Use stroke and fill attributes; avoid CSS classes
6. Use Comic Sans MS font and grayscale colors only (#000–#FFF) plus transparent fills
7. Output only the SVG markup — no prose, no code fences
8. Apply design principles (accessibility, usability, visual hierarchy) when using user comments
\end{Verbatim}

}

\vspace{1em}

\noindent
\textbf{Step 3}: Generate task flow\\
\textbf{Model}: \textsf{gemini-2.5-flash} \\
\textbf{Prompt}:
{\footnotesize
\begin{Verbatim}[breaklines,fontsize=\scriptsize]
Given a task, the SVG UI code for a sequence of screens, and the
interaction descriptions for each screen, identify the interactive
element(s) the user must engage with to proceed to the next screen.

Task:
{task}

SVG UI Codes:
{uiCodes}

Screen Interactions:
{screenInteractions}

Requirements:
- For each screen, use the interaction description to determine which
  element(s) enable transition to the next screen(s)
- Extract the complete SVG snippet for the identified element(s),
  including all attributes and nested elements if any
- Output an array of SVG snippets for each screen in order
- Be specific: avoid selecting too many elements or overly broad
  containers
- Match the interaction description to the correct UI elements

Constraints:
- Never select the entire <svg>
- Response must be a valid JSON array with no text before or after
- Do not mark the response as JSON or use code fences
- The output array must have the same length as the input uiCodes array
- If no clear interactive element is found, use an empty array for that
  screen
- Focus on elements that align with the described user interaction
  (e.g., if interaction says "click button", select the button element)
\end{Verbatim}

}

\section{Logging Details of \sys}
\label{sec:logging-details}
Below is the complete list of JavaScript events that triggered logging in \sys.
We focus on a core set of events that directly capture users' 
\emph{intentional interactions} with the GenUI tool (\eg mouse, keyboard, form, and navigation events).

\begin{itemize}
  \item \texttt{mousedown}: Triggered when a mouse button is pressed down.
  \item \texttt{mouseup}: Triggered when a mouse button is released.
  \item \texttt{click}: Fired when a mouse button is pressed and released on an element.
  \item \texttt{dblclick}: Fired when an element is clicked twice in quick succession.
  \item \texttt{mousemove}: Fired when the mouse pointer moves over an element.
  \item \texttt{contextmenu}: Fired when the user right-clicks to open a context menu.
  \item \texttt{wheel}: Fired when the mouse wheel is rotated.
  \item \texttt{dragstart}: Fired when a drag operation starts.
  \item \texttt{drag}: Fired repeatedly while an element is being dragged.
  \item \texttt{dragend}: Fired when a drag operation ends.
  \item \texttt{drop}: Fired when an element is dropped on a valid target.
  \item \texttt{keydown}: Fired when a key is pressed down.
  \item \texttt{keyup}: Fired when a key is released.
  \item \texttt{keypress}: Fired when a key producing a character value is pressed.
  \item \texttt{input}: Fired when the value of an input or textarea changes.
  \item \texttt{focus}: Fired when an element gains focus.
  \item \texttt{blur}: Fired when an element loses focus.
  \item \texttt{focusin}: Similar to \texttt{focus}, but bubbles up.
  \item \texttt{focusout}: Similar to \texttt{blur}, but bubbles up.
  \item \texttt{change}: Fired when the value of an input, select, or textarea changes.
  \item \texttt{submit}: Fired when a form is submitted.
  \item \texttt{reset}: Fired when a form is reset.
  \item \texttt{select}: Fired when text inside an input or textarea is selected.
  \item \texttt{copy}: Fired when content is copied to the clipboard.
  \item \texttt{cut}: Fired when content is cut to the clipboard.
  \item \texttt{paste}: Fired when content is pasted from the clipboard.
  \item \texttt{scroll}: Fired when an element's scrollbar is being scrolled.
  \item \texttt{resize}: Fired when the window or an element is resized.
  \item \texttt{visibilitychange}: Fired when the document's visibility changes (e.g., tab hidden).
  \item \texttt{beforeunload}: Fired before a page is unloaded, allowing save prompts.
  \item \texttt{hashchange}: Fired when the URL hash changes.
\end{itemize}

\end{document}